\documentclass[fleqn,usenatbib]{mnras}

\usepackage{newtxtext,newtxmath}

\usepackage[T1]{fontenc}
\usepackage{multirow}

\DeclareRobustCommand{\VAN}[3]{#2}
\let\VANthebibliography\thebibliography
\def\thebibliography{\DeclareRobustCommand{\VAN}[3]{##3}\VANthebibliography}

\usepackage{graphicx}	
\usepackage{amsmath}	

\newcommand{\chianti}{\textsc{chianti}\ }

\title[Modelling Stellar Irradiances I]{Modelling Stellar Irradiances I: the transition regions of FGKM stars}

\author[E. Deliporanidou, G. Del Zanna]{
E. Deliporanidou,\thanks{E-mail: ed650@cam.ac.uk}
G. Del Zanna
\\
DAMTP, Centre for Mathematical Sciences, University of Cambridge, Wilberforce Road, Cambridge, CB3 0WA, UK\\}

\date{Accepted XXX. Received YYY; in original form ZZZ}

\pubyear{2024}

\begin{document}
\label{firstpage}
\pagerange{\pageref{firstpage}--\pageref{lastpage}}
\maketitle

\begin{abstract}

We employed advanced ionisation equilibrium models that we developed in Dufresne et. al. (2024), which include charge transfer and density effects, to model UV stellar irradiances for a sample of stars. Our sample includes $\epsilon$ Eridani (K2 V), $\alpha$ Centauri A (G2 V), Procyon (F5 V) and Proxima Centauri (M5.5 Ve). We measured line fluxes from STIS datasets and used \ion{O}{iv} and \ion{O}{v} as  density diagnostics to find the formation pressure of ions in the transition region (TR) and adopted a simple Differential Emission Measure (DEM) modelling. Our findings indicate significant improvements in modelling spectral lines from anomalous ions such as Si IV, C IV and N V of the Li- and Na-like sequences, which produce the strongest lines in the UV. For example, the Si IV lines were under-predicted by a factor of five and now are within 40\% the observed fluxes. 
The improved models allow us to obtain for the first time reliable estimates of some stellar chemical abundances in the TR. We compared our results with available photospheric abundances in the literature and found no evidence for the First Ionisation Potential (FIP) effect in the TR of our stellar sample. Finally, we compared our results with the solar TR that can also be described by photospheric abundances.

\end{abstract}


\section{Introduction}
\label{sec:Intro}

This paper is the first in a series where we study XUV stellar irradiances,
aiming to improve the atomic data and provide accurate modelling.
Measuring XUV stellar irradiances is  important for modelling exoplanet atmospheres \citep[see, e.g.][]{linsky_host_2019}.
The EUV has a particular importance, but 
 very few observations of nearby stars are available from 
low-resolution EUVE spectra. We will discuss the modelling of 
the EUV in future papers. 

In the UV, there are plenty of observations from e.g. the 
Space Telescope Imaging Spectrograph (STIS), with
primarily emission from the chromospheres and transition regions (TR). It is now well known that the TR of the Sun and stars 
have little variability with the activity cycles
\citep[see, e.g.][]{del_zanna_euv_2015, ayres_far-ultraviolet_2015}. 
However, the modelling of the strongest TR UV lines is notoriously
difficult: zero density ionisation equilibrium (the so-called coronal approximation, the standard in astrophysics),
in combination with simple emission measure modelling have failed 
to accurately predict the observed intensities of lines from the 
`anomalous' ions, those belonging to the Li- and Na-like sequences, such as \ion{Si}{iv}, \ion{C}{iv} and \ion{N}{v}. Such ions produce the strongest emission lines in the UV, which are under-predicted by significant factors (2--5) compared to those from other ions forming at similar temperatures. 
 This problem was observed on the Sun for a long time (see the review by \cite{del_zanna_solar_2018}).  \citet{del_zanna_spectroscopic_2002} combined EUV and UV observational data to show for the first time that these anomalous ions are also underpredicted by large factors in stellar atmospheres of active stars.

 \citet{sim_modelling_2005} developed an approximate ionisation equilibrium by including density and charge transfer effects for a few carbon and silicon ions and applied it, together with simple 1D emission measure analyses, to UV observations of $\epsilon$ Eridani, which is a K2 V dwarf star. They claimed to have resolved the main discrepancies in the anomalous ions, although close inspection of their results still indicate 
 large deviations for several lines. 
 On the other hand,  \cite{judge_failure_1995} also developed improved
 ionisation equilibrium models similar to those of 
 \citet{sim_modelling_2005}, but did not find significant improvements
 in the anomalous ions when applying an emission measure modelling to solar irradiances. Their conclusion was that other physical effects are at play.

We also developed advanced ionisation equilibrium models that include density dependent and charge transfer effects with updated atomic rates,
and recently made them available to the community 
via \chianti version 11 \citep{dufresne_chianti_2024}.
We found significant improvements when comparing observed and predicted radiances for a solar plage observation in the UV, 
considering the same lines regularly observed by HST/STIS. The main differences between the advanced models and the coronal approximation used in the previous \chianti versions \citep{dere_chianti_1997, del_zanna_chiantiatomic_2021}, is that the anomalous ions have increased intensities by factors of 2-5 and the formation temperatures of all TR lines are shifted towards lower values.

The first aim of this paper is therefore to quantify 
 improvements in modelling the lines from anomalous ions in stellar transition regions. 
 We adopt a simple 1D DEM modelling, as common in the literature,
 although we point out that this is an unphysical representation 
of the whole atmosphere of a star. 
The second aim is to use these advanced models to derive,  
to our knowledge for the first time, relative chemical 
abundances in the transition regions of a sample of stars.

The present paper  focuses on STIS UV observations  of the transitions regions of four  stars: $\epsilon$ Eridani (K2 V), $\alpha$ Centauri A (G2 V), Procyon (F5 V), and Proxima Centauri (M5.5 Ve). 
Our stellar sample was chosen to investigate the First Ionisation Potential (FIP) effect  in stellar TR. This effect, first seen in the active solar corona \citep{pottasch_lower_1963}, is characterised by an overabundance of low-FIP elements ($<10$ eV), such as Fe, Mg and Si compared to high-FIP elements ($>10$ eV), such as  C,N, O. \citet{telleschi_coronal_2005} note that more active, younger stars have a different FIP effect compared to older, less active stars. 
For a review of the solar literature, where several 
results turned out to be incorrect, see \cite{del_zanna_solar_2018}. The photospheric composition of the solar TR was noticed in an early study from \citet{laming_stellar_1995}, and this has been reaffirmed in more recent reviews, such as \cite{del_zanna_solar_2018}. As solar irradiances in the transition region vary little (20-30\%) 
with the  solar cycle \citep{del_zanna_euv_2015}, the active 
Sun as a star also has near photospheric abundances in the TR.

\citet{laming_unified_2004} developed theoretical models to explain the physical mechanisms behind the FIP effect, proposing that the enhancement of low-FIP elements occurs due to the ponderomotive force acting on ionised particles in the chromosphere and TR. 
An explanation for the inverse FIP effects was also put forward.
Therefore, the FIP and inverse FIP effects are not just an 
annoying feature when modelling stellar irradiances, but 
could in principle be turned into a significant diagnostic tool of stellar magnetic activity and atmospheric dynamics \citep{laming_first_2017}. 

EUV and X-ray observations of the hot plasma in stellar coronae 
have shown a large variety of results,  with both FIP and 
the opposite, Inverse IFIP (IFIP) effects 
present among active stars and with variations with stellar activity, see e.g. the reviews by \cite{laming_fip_2015,seli_extending_2022}. To study further the FIP dependence on the stellar activity, \citet{wood_resolving_2010} compared EM distributions and coronal abundances of seven moderately active dwarfs of different stellar types. They found that the time-dependent interpretation can only be tested if there are multiple observations of the same star at different times. They did find however, a strong correlation of FIP bias with the spectral type, with earlier type stars having solar-like FIP effects and the later stars having inverse FIP effects.

Earlier studies suffered from many limitations. 
A significant limitation  was that stellar coronal abundances were 
often measured relative to the solar photospheric 
abundances, whenever stellar  measurements were not available.
Within the past two decades, a large number of surveys have 
now provided many stellar photospheric abundances, so the FIP
effect can now be properly studied. 
Another limitation of earlier  studies is related to 
the significant changes in the solar photospheric abundances 
of key high-FIP elements such as C,N,O.

On the basis of the solar results, we would expect to find 
chemical abundances close to the photospheric ones in the 
transition regions of our stellar sample,  
as stellar TR also tend to vary little 
\citep[see, e.g.][]{ayres_far-ultraviolet_2015}.
To our knowledge,
such measurements have not been previously attempted. 
The advanced models do give us the opportunity to 
provide such measurements, although for only a few elements.

\begin{table*}
    \centering
    \caption{Stellar Parameters and Observational Data for the four active dwarf stars}
    \begin{tabular}{c|c|c|c|c|c|c|c}
    \hline
   \textbf{Star name} & \textbf{Type} & \textbf{STIS dataset} & \textbf{Date} & \textbf{$T_{\text{eff}}$ [K]} & \textbf{Distances [pc]} & \textbf{Radius [R/R$_{\odot}$]} & \textbf{Pressure [$\text{cm}^{-3} \text{K}$}] \\
    \hline
    $\epsilon$ Eridani & K2 V & O55P01030 & 2000-03-17 & 5180 & 3.22 & 0.74 & $1\times10^{16}$ \\
    A Centauri A & G2 V & O53P01030 & 1999-02-12 & 5770 & 1.34 & 1.23 & $5\times10^{14}$ \\
    Procyon &  F5 1V & OBKK13040 & 2011-04-05 & 6704 & 3.51 & 2.05 & $1\times10^{16}$\\
    Proxima Centauri & M5.5 Ve & O5EO01010 & 2000-05-08 & 3050 & 1.29 & 0.15 & $1\times10^{16}$ \\ 
    \hline
    \end{tabular}
    \label{tab:stellar_details}
\end{table*}

We  detail the methods used in our analysis in Section~\ref{Methods}. In Section~\ref{methodology_data} we describe the observational data, the chosen stellar sample, and the relevant literature, including the available photospheric abundances.  In section~\ref{Results} we present our results that include improved DEM modelling, enhanced line ratios and estimates for the stellar TR abundances. In section~\ref{Discussion} we compare our results with existing photospheric abundance measurements, assessing the FIP effect within our stellar sample and finally compare our findings with solar TR irradiances in the EUV.

\section{Methods}
\label{Methods}

To avoid dealing with radiative transfer effects for this initial 
study, we have chosen 
 optically thin lines and did not consider singly ionised atoms, also for the following reasons: as shown in \citet{dufresne_influence_2021, dufresne_benchmark_2023}, singly ionised atoms are also 
 affected by photoionisation which makes the modelling more complex. 
We have assessed their atomic data and have found that they are often 
not very accurate. We will discuss these issues and provide 
improvements in follow-up papers. Our calculations included photoexcitation 
of the disk radiation (assuming a black-body with the $T_{\rm eff}$ of each star), which also affects some low charge states.

We measured line fluxes from STIS datasets (see Table~\ref{tab:stellar_details}) and utilised \ion{O}{iv}, \ion{O}{v} lines as a density diagnostic. We assumed a constant pressure for ions in the stellar TR, as commonly adopted in the literature (note that radiative transfer 
calculations for a static atmosphere do indicate constant
pressure in the TR). We finally used a pressure of $10^{16}$ cm$^{-3}$ K for $\epsilon$ Eridani, Proxima Centauri, and Procyon and a pressure of $5 \times 10^{14}$ cm$^{-3}$ K for $\alpha$ Centauri A.

We eventually evaluated the improvements in modelling anomalous ions and derived reliable estimates of stellar abundances via a DEM modelling using the CHIANTI\_DEM routine and the default MPFIT option.

Once we obtain the fluxes we can either obtain a volume or a column
emission measure. Neglecting interstellar absorption, 
the flux in an optically thin line  of wavelength $\lambda_{ij}$ 
(frequency $\nu_{ji}$) from a star at a distance $d$ is:
\begin{equation}
F(\lambda_{ji}) = {{h \nu_{ji}} \over 4 \pi d^2 }~ 
\int_V ~  N_j A_{ji} ~dV \quad
[\mathrm{erg\ cm^{-2}\ s^{-1}}]
\end{equation}
where $N_j$ is the population of the upper state, $V$
is the emitting volume, and $A_{ji}$ is Einstein's A-value.
The relation is usually written as:
\begin{equation}
F(\lambda_{ji}) = {{1 \over d^2 } \, \int_V ~ Ab ~ G(T,N_{\rm e}) \, N_{\rm e} N_{\rm H} ~dV }
\end{equation}
where $Ab$ is the elemental abundance, 

$N_{\rm e} N_{\rm H}$ the electron
and neutral hydrogen number densities, and  $T$ is the electron
temperature.  
$G$ is the contribution function of the line calculated
by the CHIANTI programs, in erg cm$^{3}$ s$^{-1}$ sr$^{-1}$:
 \begin{equation}
G(N_{\rm e},T) =  A_{ji} ~{h \nu_{ij} \over{4\pi } } 
{N_j(Z^{+r})\over (Z^{+r})}   { N(Z^{+r}) \over N(Z)}   
{1 \over N_{\rm e}}
\end{equation}
where  ${N_j(Z^{+r}) / N(Z^{+r})}$
is the population of the upper level $j$, relative to the total ion 
population,
${N(Z^{+r}) / N(Z)}$ is the ion abundance, calculated with the 
advanced v.11 model.

Assuming that there is a single relation between the 
densities and the electron temperature, one defines a volume 
DEM as 
$$ DEM_v = N_{\rm e} N_{\rm H} ~dV/dT \quad [{\rm cm}^{-3}\ {\rm K}^{-1}] . $$
Our choice here is instead to 
 assume spherical symmetry, i.e. 
 that the stellar emission comes from $dV=4 \pi r^2 dh$, so we have
\begin{equation}
F(\lambda_{ji}) = {4 \pi \, r^2 \over d^2 }~ 
\int_T ~ Ab \, G(T,N_{\rm e}) \, DEM(T) \, dT \quad 
\end{equation}
by assuming that a single-valued column DEM exists: 
$$ DEM(T) = N_{\rm e} N_{\rm H} ~dh/dT \quad [{\rm cm}^{-5}\ {\rm K}^{-1}] .$$ 
In this way, we can compare the DEM values among the different 
stars and the Sun, as we know their distance and radius.  Note that in the literature the emission measures, which are the integral in temperature bins
of the DEM are often presented. Also, often the DEM is defined as $ DEM(T) = N_{\rm e}^2 ~dh/dT $.

We used the \citet{fontenla_far-_2014} facula model atmosphere for all stars to calculate charge transfer for the ion balances, as its pressure through the TR aligns well with the pressures derived from the observations.

When reviewing the literature on photospheric measurements, 
we encountered several issues. 
First, different methods and choices of atoms / lines normally 
produce different results. Second, significant variations (30--50\%)
are found in the literature results.  Third, since measurements are normally reported as relative to solar abundances, the uncertainty in the solar
values is an additional factor. 
We have chosen as reference the latest
solar photospheric values by \citet{asplund_chemical_2021}, noting that
the abundances of elements such as C,N,O are significantly (more than 30\%)
different from the values used in the earlier literature. 
Nevertheless, we point out that the FIP bias has values that typically range
between 2 and 4, so the cited uncertainties in the photospheric 
abundances do not affect the main results shown here. 

To address these challenges, we carefully analysed the photospheric abundances for each star via our DEM analysis. The DEM provides a measure of the distribution of emitting plasma at different temperatures in the stellar atmosphere but can also serve as a diagnostic tool for measuring elemental abundances using spectral lines \citep{del_zanna_solar_2018}.

We specifically chose the best fit of the DEM while varying elemental abundances for all stars. By applying the advanced models, we investigated discrepancies between the various abundance values reported in the literature. Given the improvements observed in the transition region modelling of the Sun \citep{dufresne_chianti_2024}, we trusted these models to provide a measure of the relative abundance values for each star. This selection was also based on achieving the best predicted-to-observed flux ratio.

\section{Data}
\label{methodology_data}

\subsection{Observational Data}

The datasets for all four stars were obtained from STIS, using either the E140H or E140M gratings,  and within the primary wavelength range of $1100<\lambda<1700$. We used the recently (December 2023) reprocessed 
STIS data from the MAST website, which provides an uncertainty measurement for each pixel in the data. 

Our selection was based on the longest exposure times and the dates, choosing datasets closer to the launch date of the instrument to minimise potential degradation effects.
Additionally, an important criterion was the availability of Chandra LETG datasets for the same observational dates to extend our analysis to X-Ray spectra in future papers and further investigate the FIP effect in the coronae of these stars. 

Details such as  dataset names and observation times, along with essential parameters for each stellar object, are provided in Table~\ref{tab:stellar_details}. Our analysis showed minor differences between the E140H and E140M gratings.

 The line fluxes were obtained with the {\it cfit} package, developed  by S.V.Haughan for SOHO, and custom-written programs. 
 We applied  a rebinning factor of 2 to enhance the signal-to-noise ratio, and removed a background.
 The fitting assumed Gaussian line profiles, which is very accurate for most cases. We calculated the effective temperature (T$_p$) as the average of the temperature, weighted by the product of the line contribution function, G(T), with the DEM:

 \begin{equation}
{\rm log}\, T_{\rm p} = { \int G{\left({T}\right)}~
DEM{\left({T}\right)} ~{\rm log}~ T~dT \over
{\int G{\left({T}\right)}~DEM{\left({T}\right)}~dT} } \quad .
     \label{T_eff}
 \end{equation}
 
 We have also compared the line fluxes with those obtained by summing the 
 signal over the entire line profile, removing a background, along 
 the lines of \citet{ayres_far-ultraviolet_2015}. 
 We found minor differences, of the order of a few percent, between 
 the two methods, for the stronger lines. 
 We have also compared our results to any available in the literature,
 finding small differences, typically within 10\%.

In addition, we closely looked at  strong lines such as the \ion{C}{iii} multiplet lines to make sure 
that the wavelengths of the lines were at rest for the fitting (as we are not interested in measuring Doppler shifts).
 We found that A Cen A was blue-shifted by $\lambda=$0.1 \AA, and $\epsilon$ Eridani was blue-shifted by $\lambda=$0.8 \AA.

\subsection{Atomic Data and line selection}

We have used data from the latest \chianti v.11 release, \citep{dufresne_chianti_2024}. For UV emissions, interstellar absorption is a significant problem below 912 \AA, where the ISM is opaque due to strong absorption by neutral hydrogen. Above 912 \AA, while there is still absorption (notably around specific lines like the Lyman-alpha 1216~\AA\ line), the ISM becomes more transparent, allowing for more effective observations of UV emissions from astronomical sources.

In addition, there are blended or self-absorbed lines like the  1526.7~\AA\ \ion{Si}{ii}, 1404~\AA\ \ion{Si}{iv} and \ion{O}{iv} as well as the 1334.53~\AA\ and 1335.7~\AA\ \ion{C}{ii}. We also found that the \ion{C}{ii} lines were completely non-Gaussian. When fitting the lines, we also included a few nearby neutral lines around the 1401 Å \ion{O}{iv} line, as observed in the IRIS spectra, to increase the accuracy of the obtained flux values. However, we did not include any neutral ions in the actual DEM model. 

To measure the electron densities, we considered all 
the \ion{O}{iv} multiplet lines, but used 
the ratio of the $\lambda=$ 1401~\AA\  and $\lambda=$ 1407~\AA\  \ion{O}{iv} lines, as the 1404~\AA\ is blended, and the
1399~\AA\ is often very weak and tends to indicate lower 
densities, when used in conjunction with the 1401~\AA\  line,
as we found in the solar case \citep{rao_path-lengths_2022}.

We have also considered the
 \ion{O}{v} $1218$ vs. $1371$~\AA\ line ratio, although 
 it has some temperature dependency.
 To obtain the pressures, we considered the effective
 temperature of formation of these ions with the 
 advanced equilibrium models and the emission measure distributions.

\subsection{Stellar Sample}
\label{Stellar Sample}

\subsubsection{$\epsilon$ Eri}

$\epsilon$ Eridani (K2 V) is a nearby, young, and active dwarf star located at approximately 3.22 pc away. $\epsilon$ Eridani's UV spectrum has been extensively analysed using the STIS high-resolution spectra, for understanding the ionisation states and abundances of various elements in its atmosphere. 

\citet{jordan_chromospheres_1987} conducted early UV studies using the International Ultraviolet Explorer (IUE) and found strong emission lines indicative of significant chromospheric heating. \citet{sim_modelling_2005} addressed the issue of the anomalous ions by presenting new ionisation balance calculations for carbon and silicon, with an approximate treatment of dielectronic recombination (DR) suppression and charge transfer (CT). \citet{ness_corona_2007} focused on the 
coronal properties using X-ray observations. They found no difference between photospheric and coronal abundances for a few elements. 
 
  We reviewed the literature and found that 
  \citet{zhao_chemical_2002} and \citet{allende_prieto_s4n_2004} provided accurate metallicities for $\epsilon$ Eridani. 
As in many other cases we have looked at, significant variations 
in the results are found. Also, as usual, measurements are reported as 
relative to the solar abundances. 
The resulting   $\epsilon$ Eri photospheric abundances are 
  shown in Table~\ref{tab:eridani abundances}. By assessing them, we believe that the \citet{zhao_chemical_2002} values are the most reliable to use for our  analysis. 

\begin{table}
    \centering
    \caption{$\epsilon$ Eridani photospheric abundances available in the literature by \citet{zhao_chemical_2002, allende_prieto_s4n_2004}, applied to solar photospheric values by \citet{asplund_chemical_2021}.}
    \begin{tabular}{c|c|c|c}
        \hline
          & \shortstack{\textbf{Zhao} \\ \textbf{2002}} &  \shortstack{\textbf{Allende} \\ \textbf{2004}}  &  \shortstack{\textbf{This} \\ \textbf{work}} \\
        \hline
          H & 12.00 & 12.00 & 12.00 \\
         C & 8.51 & 8.22 & 8.51 \\
         N & 7.79 & - & 7.85 \\
         O & 8.65 & 8.65 & 8.68 \\
         Mg & 7.53 & 7.52 & 7.57 \\
         Si & 7.47 & 7.50 & 7.53 \\
         S & 7.23 & - & 7.18\\
         Fe & 7.34 & 7.40 & 7.51\\
        \hline
        $T_{eff}$ [K] & 5104 & -- & 5104 \\
        \hline
    \end{tabular}
    \label{tab:eridani abundances}
\end{table}

\citet{laming_stellar_1996} conducted the first coronal abundances analysis of $\epsilon$ Eridani by analysing EUV spectra, and found some evidence of a solar-like FIP effect. \citet{sanz-forcada_coronal_2004} determined the coronal abundances of four stars including $\epsilon$ Eridani and reviewed photospheric abundances to investigate the FIP effect. Contrary with \citet{laming_stellar_1996}, they found a mild FIP-like effect which affects only 2 elements (Ca and Ni), showing a gradual decrease in coronal abundance for elements with increasing FIP.  In addition \citet{wood_resolving_2010} used $\epsilon$ Eri for a comparison of the EM distributions of seven active dwarfs, based on results from their previous studies \citep{wood_coronal_2006}. They compared coronal to assumed stellar photospheric abundances and they found a weak FIP effect for $\epsilon$ Eri, similar to a few more other active stars.

We calculated the pressure of $\epsilon$ Eridani, using the ratio of the $\lambda=$ 1401~\AA\  and $\lambda=$ 1407~\AA\ \ion{O}{iv} fluxes as a density diagnostic and found a pressure of $P_{\rm e}=1\times10^{16} \rm{cm} ^{-3}$ K.

\subsubsection{$\alpha$ Centauri A}

$\alpha$ Centauri A (also known as  Alpha Cen A) is a G2V-type main-sequence star, one of the closest stellar neighbours to our solar system at a distance of approximately 1.34 pc. It forms part of the Alpha Centauri triple star system, which includes Alpha Centauri B and Proxima Centauri. $\alpha$ Centauri A is a solar like star but slightly larger and more luminous than the Sun, with a mass of about 1.1 solar masses and a radius of roughly 1.2 R$_{\odot}$. Despite these differences, many researchers have noted the STIS A Cen A spectrum to be very close 
to that of the 
Sun as a star \citep[cf.][]{pagano_hststis_2004}.

\begin{table}
    \centering
    \caption{A Centauri A photospheric abundances obtained from the literature, Morel et. al. 2018, Edvardsson et. al. 1993, Buder et. al. 2021 (although the N and O abundances are approximated). For the Morel et. al. 2018, we used the after GCE abundances.}
    \begin{tabular}{c|c|c|c|c}
        \hline
         & \shortstack{\textbf{Morel} \\ \textbf{2018}} & \shortstack{\textbf{Edvardsson} \\ \textbf{1993}} & \shortstack{\textbf{Buder} \\ \textbf{2021}} & \shortstack{\textbf{Allende} \\ \textbf{2004}} \\
        \hline
         H & 12.00 & 12.00 & 12.00 & 12.00 \\
         C & 8.69 & 8.62 & 8.56 & 8.62 \\
         N & 8.09 & - & 7.96 & 7.83  \\
         O & 8.88 & 8.69 & 8.82 & 8.94 \\
         Na & 6.57 & 6.24 & 6.71 & 6.24 \\
         Mg & 7.85 & 7.88 & 8.05 & 7.99 \\
         Al & 6.73 & 6.45 & 8.87 & 6.45 \\
         Si & 7.77 & 7.51 & 7.66 & 7.83  \\
         S & 7.38 & - & 7.25 & 7.12  \\
         Ca & 6.56 & 6.50 & 6.75 & 6.74  \\
         Fe & 7.74 & 7.65 & 7.75 & 7.62 \\
         Ni & 6.51 & 6.42 & 6.60 & 6.42  \\
         \hline
         $T_{eff}$ [K] & 5795 & 5720 & 5271 & 5519 \\
        \hline
    \end{tabular}
    \label{tab:A Cen Aabundances}
\end{table}

\citet{pagano_hststis_2004} analysed the UV spectrum of Alpha Centauri A, providing measurements of emission line fluxes. They highlighted similarities and differences in UV spectra between Alpha Centauri A and the Sun, noting comparable activity levels but variations in specific line intensities due to differences in magnetic activity and heating mechanisms. \citet{ayres_far-ultraviolet_2015} used high-resolution E140M echelle spectrograms from STIS and studied the FUV emissions of A Cen A. \citet{judge_comparison_2004} wanted to determine how A Cen A serves as a proxy for the Sun under intermediate activity conditions, using datasets from STIS. They found similarities in chromospheric line intensities between A Cen A and the Sun under intermediate activity conditions, but slight differences in the TR and coronal emissions. Upon evaluating the fluxes available in the literature and the values we obtained by fitting the lines, we found agreement within 10 $\%$ between the published fluxes of \citet{pagano_hststis_2004, ayres_far-ultraviolet_2015, judge_comparison_2004} and ours, if scaled accordingly with the provided scaling factors.

We calculated the pressure of A Cen A using both \ion{O}{iv} and \ion{O}{v} line fluxes and found that the star is in the low-density limit, hence we used a pressure of P$_{\rm e}=5\times10^{14}$ cm$^{-3}$ K.

Upon reviewing the literature on photospheric abundances for Alpha Centauri A, we noted that many researchers have used solar photospheric values. We did find specific measured values reported by \citet{morel_chemical_2018}, \citet{edvardsson_chemical_1993}, \citet{buder_vizier_2021}, and \citet{allende_prieto_s4n_2004}, which we present on Table~\ref{tab:A Cen Aabundances}, scaled to the solar photospheric values by \citet{asplund_chemical_2021}. Among these, the best results from our DEM analysis were obtained using the photospheric abundances reported by \citet{morel_chemical_2018}.

\citet{raassen_chandra-letgs_2003} analysed Chandra LETG X-ray data for A Cen A and observed that the low FIP elements were more abundant in the star's corona, indicating the presence of an FIP effect. This confirms the initial findings of \citet{drake_stellar_1997}, who found that the low-FIP elements are overabundant relative to the high-FIP elements by a factor of about 2.

\subsubsection{Procyon}

Procyon (Alpha Canis Minoris) is one of the brightest stars in the night sky and is a binary star system consisting of Procyon A, a F5 IV-V subgiant or main-sequence star, and Procyon B, a faint white dwarf. The system is relatively close to Earth, at a distance of approximately 3.51 pc.

When reviewing the literature for photospheric abundances for Procyon, we found measured values by \citet{allende_prieto_s4n_2004, buder_galah_2021, edvardsson_chemical_1993, drake_stellar_1995} which we list in table~\ref{tab:procyon_abundances}. Among these, we found very good results when using the abundances by \citet{drake_stellar_1995}, which are solar photospheric abundances from \citet{anders_abundances_1989}, except for iron for which the researchers provide a different value. We list these in table~\ref{tab:procyon_abundances} under "This Work". In the same table, we also provide the solar photospheric abundances from \citet{asplund_chemical_2021} to highlight the differences, as the solar photospheric composition has changed significantly since 1989. However, we chose to proceed with the abundances from \citet{drake_stellar_1995} 

Contrary to the solar case, studies by \citet{raassen_high-resolution_2002} and \citet{schmitt_coronal_2004} found a less significant FIP effect for Procyon. \cite{sanz-forcada_coronal_2004} found agreement between photospheric and coronal abundances with a relative underabundance of O in the corona. They also found an indicative abundance trend of a mild FIP bias which increases with temperature in the corona of Procyon, but couldn't confirm the hypothesis due to uncertainties. \citet{drake_stellar_1995} also found that the corona of Procyon did not exhibit clear evidence of a solar-like FIP effect and that the observed abundances were all, to within the observational uncertainties, identical to Procyon's photospheric abundances.

We calculated the pressure of Procyon using the ratio of the 1407~\AA\ and 1401~\AA\ \ion{O}{iv} lines, finding a high pressure of $P_{\rm e}=10^{16} \text{cm}^{-3}$ K. Under the correct pressure, we would expect the ratio of the calculated-to-observed fluxes of the 1218~\AA\ and 1371~\AA\ \ion{O}{v} lines to be consistent. However, the flux ratios of these lines did not agree at $P_{\rm e}=$\(10^{16} \, \text{cm}^{-3} \, \text{K}
\). Using the 1218~\AA\ and 1371~\AA\ \ion{O}{v} lines as density diagnostics, we found a slightly higher pressure of $P_{\rm e}=$ \(5 \times 10^{16} \, \text{cm}^{-3} \, \text{K}
\), although this did not yield good calculated/observed flux ratios for the other ions. Given that Procyon is a quiet star, a high pressure is unexpected, leading us to adopt the initial measurement of $P_{\rm e}=$ \(10^{16} \, \text{cm}^{-3} \, \text{K}
\) based on the \ion{O}{iv} lines.

\begin{table}
    \centering
    \caption{Procyon photospheric abundances available in the literature measured by \citet{buder_vizier_2021, edvardsson_chemical_1993,  allende_prieto_s4n_2004} and solar photospheric abundances by \citet{asplund_chemical_2021} for reference. "This work" consists of abundances used from \citet{drake_stellar_1995} except for C and N that are taken from \citet{anders_abundances_1989}.}
    \begin{tabular}{c|c|c|c|c|c|c}
        \hline
          & \shortstack{\textbf{Buder} \\ \textbf{2021}}  & \shortstack{\textbf{Edvardsson} \\ \textbf{1993}} & \shortstack{\textbf{This} \\ \textbf{Work}}  &  
          \shortstack{\textbf{Asplund} \\ \textbf{2021}}  &  
          \shortstack{\textbf{Allende} \\ \textbf{2004}}  \\
        \hline
         H & 12.00 & 12.00 & 12.00 & 12.00 & 12.00 \\
         C & 8.44 & - & 8.56 & 8.46 & 8.51 \\
         N & 7.84 & - & 8.05 & 7.83 & - \\
         O & 8.70 & 8.64 & 8.93 & 8.69 & 8.67 \\
         Na & 6.25 & 6.28 & - & 6.22 & - \\
         Mg & 7.61 &  7.67 & 7.58 & 7.55 & 7.54 \\
         Al & 6.45 & - & - & 6.43 & - \\
         Si & 7.52 & 7.52 & 7.55 & 7.51 & 7.44 \\
         S & 7.13 & - & 7.21 & 7.12 & - \\
         Ca & 6.34 & 6.30 & - & 6.30 & 6.70 \\
         Fe & 7.51 & 7.58 & 7.51 & 7.46 & 7.49 \\
         Ni & 6.23 & 6.24 & 6.25 & 6.20 & 6.27 \\
        \hline
        $T_{eff}$ [K] & 6468 & 6704 & 6650 & - & 6677 \\
        \hline
    \end{tabular}
    \label{tab:procyon_abundances}
\end{table}

\subsubsection{Proxima Centauri}

Proxima Centauri is a red dwarf of spectral class M 5.5, and is part of the Centauri triple stellar system. It is known to have a high level of magnetic activity, which can result in intense flares. It is the closest star to our solar system and the closest known exoplanetary host star candidate \citep{lalitha_proxima_2020}. 

Despite finding extensive literature on coronal abundances for Proxima Centauri, actual measurements of its photospheric abundances are lacking. Consequently, we decided to adopt the photospheric abundances of Alpha Cen A from \citet{morel_chemical_2018}, based on the assumption that Proxima Centauri, as a member of the Alpha Centauri triple system, shares a similar atmospheric composition due to its formation from the same primordial nebula cloud.

\citet{fuhrmeister_multi-wavelength_2011} found an inverse FIP pattern during the quiescent state for Proxima Centauri, which aligns with \citet{lalitha_proxima_2020} who discussed that an anti-FIP effect is expected during the quiescent emission of moderately active M dwarf stars. \citet{fuhrmeister_astronomy_2022} anticipated an inverse-FIP abundance pattern for Proxima Centauri based on its stellar type. However, their observations were inconclusive regarding the presence of an FIP or inverse-FIP effect, noting that the abundance pattern was relatively close to solar photospheric values, which could be attributed to the quiescent state of Proxima Centauri during the observations. Similarly, \citet{gudel_flares_2004} identified a rather flat abundance pattern that was also close to solar photospheric values.

We calculated the pressure of Proxima Centauri using the \ion{O}{iv} lines and obtained a range of pressures from \(P_{\rm e} = 10^{16}\ \text{cm}^{-3} \, \text{K}
\) to \(P_{\rm e} = 5 \times 10^{16}\ \text{cm}^{-3} \, \text{K}
\), ultimately selecting \(P_{\rm e} = 10^{16}\ \text{cm}^{-3} \, \text{K}
\) as it provided the most consistent results with our observations.

\section{Results}
\label{Results}

\begin{table*}
    \centering
    \caption{Numerical results of the DEM analysis for all four stars. $F_{\text{obs}}$ is the measured flux in $10^{-14}$ erg cm$^{-2}$ s$^{-1}$ . $\log T_{\text{p}}$ is the effective temperature is [K] and $\frac{F_{\text{calc}}}{F_{\text{obs}}}$ is the ratio of the calculated/predicted to the observed fluxes. }
    \begin{tabular}{c|c|c|c|c|c|c|c|c|c|c|c|c|c}
        \hline
        \hline
        \multirow{2}{*}{Wavelength (\AA)} & \multirow{2}{*}{Ion} & \multicolumn{3}{c|}{$\epsilon$ Eridani} & \multicolumn{3}{c|}{$\alpha$ Centauri A} & \multicolumn{3}{c|}{Procyon} & \multicolumn{3}{c}{Proxima Centauri} \\
        \cline{3-5} \cline{6-8} \cline{9-11} \cline{12-14}
        & & $\log T_{\text{p}}$ & $F_{\text{obs}}$ & $\frac{F_{\text{calc}}}{F_{\text{obs}}}$ & $\log T_{\text{p}}$ & $F_{\text{obs}}$ & $\frac{F_{\text{calc}}}{F_{\text{obs}}}$ & $\log T_{\text{p}}$ & $F_{\text{obs}}$ & $\frac{F_{\text{calc}}}{F_{\text{obs}}}$ & $\log T_{\text{p}}$ & $F_{\text{obs}}$ & $\frac{F_{\text{calc}}}{F_{\text{obs}}}$ \\
        \hline
        1190.42 & \ion{Si}{ii} & 4.05 & 1.09 & 0.70 & 4.13 & 3.24 & 0.39 & 4.17 & 6.38 & 0.49 & 4.16 & 0.01 & 1.14 \\
        1197.39 & \ion{Si}{ii} & 4.06 & 0.92 & 0.92 & 4.13 & 1.59 & 0.86 & 4.19 & 3.95 & 0.85 & 4.19 & 0.05 & 0.33 \\
        1526.71 & \ion{Si}{ii} & - & - & - & 4.07 & 24.5 & 0.86 & 4.11 & 21.0 & 1.28 & 4.09 & 0.10 & 0.78 \\
        1259.53 & \ion{S}{ii} & 4.08 & 1.25 & 1.67 & - & - & - & 4.31 & 14.2 & 0.55 & 4.23 & 0.06 & 0.96 \\
        1253.81 & \ion{S}{ii} & - & - & - & 4.27 & 2.07 & 0.73 & 4.35 & 9.12 & 0.63 & 4.27 & 0.03 & 1.62 \\
        1335.71 & \ion{C}{ii} & 4.27 & 18.6 & 1.30 & 4.35 & 739 & 0.99 & 4.41 & 317 & 0.60 & 4.43 & 1.13 & 1.35 \\
        1334.53 & \ion{C}{ii} & 4.27 & 45.7 & 1.05 & 4.35 & 17.1 & 0.84 & 4.40 & 499 & 0.74 & 4.42 & 2.58 & 1.15 \\
        1206.51 & \ion{Si}{iii} & 4.54 & 33.1 & 0.82 & 4.56 & 152 & 0.77 & 4.57 & 476 & 1.11 & - & - & - \\
        1323.97 & \ion{C}{ii} & 4.62 & 0.12 & 1.01 & 4.60 & 0.69 & 0.74 & - & - & - & 4.74 & 0.04 & 0.38 \\
        1298.95 & \ion{Si}{iii} & 4.63 & 1.67 & 0.97 & 4.63 & 3.96 & 0.96 & 4.62 & 18.0 & 1.92 & 4.65 & 0.22 & 1.22 \\ 
        1294.54 & \ion{Si}{iii} & 4.63 & 0.59 & 0.79 & 4.64 & 1.22 & 0.91 & - & - & - & 4.65 & 0.13 & 0.59 \\
        1200.97 & \ion{S}{iii} & 4.64 & 0.64 & 1.10 & 4.66 & 1.96 & 1.36 & 4.63 & 14.3 & 1.00 & 4.65 & 0.12 & 0.86 \\
        1194.02 & \ion{S}{iii} & 4.65 & 0.69 & 0.62 & 4.66 & 1.11 & 1.51 & 4.64 & 7.61 & 1.17 & 4.65 & 0.09 & 0.74 \\
        1296.61 & \ion{S}{iv} & 4.70 & 0.26 & 1.42 & 4.69 & 0.99 & 0.88 & 4.69 & 4.97 & 1.60 & - & - & -\\
        1175.71 & \ion{C}{iii} & 4.77 & 10.9 & 1.28 & 4.75 & 2.81 & 1.43 & 4.76 & 128 & 1.30 & 4.74 & 1.62 & 1.27 \\
        1175.99 & \ion{C}{iii} & 4.77 & 4.85 & 0.76 & 4.75 & 92.0 & 1.07 & 4.77 & 49.9 & 0.89 & 4.74 & 0.63 & 0.88 \\
        1176.37 & \ion{C}{iii} & 4.77 & 4.00 & 1.16 & 4.75 & 12.8 & 1.04 & 4.77 & 68.7 & 0.81 & 4.75 & 80.8 & 0.85 \\
        1175.26 & \ion{C}{iii} & 4.77 & 3.86 & 0.97 & 4.75 & 10.2 & 1.04 & 4.77 & 38.3 & 1.16 & 4.74 & 0.77 & 0.72 \\
        1174.93 & \ion{C}{iii} & 4.77 & 3.91 & 1.19 & 4.76 & 12.5 & 1.07 & 4.77 & 84.9 & 0.65 & 4.74 & 0.90 & 0.76 \\
        1175.59 & \ion{C}{iii} & 4.77 & 2.89 & 0.97 & 4.75 & 12.7 & 0.64 & 4.76 & 65.0 & 0.52 & 4.75 & 0.74 & 0.56 \\
        1402.77 & \ion{Si}{iv} & 4.82 & 10.6 & 0.63 & 4.79 & 41.0 & 0.54 & 4.79 & 139 & 0.80 & 4.80 & 0.96 & 1.02 \\
        1393.76 & \ion{Si}{iv} & 4.82 & 19.9 & 0.67 & 4.79 & 79.4 & 0.55 & 4.80 & 246 & 0.90 & 4.80 & 1.64 & 1.19 \\
        1666.16 & \ion{O}{iii} & 4.85 & 0.58 & 0.73 & 4.81 & 11.1 & 0.41 & 4.83 & 25.2 & 0.52 & - & - & - \\
        1660.80 & \ion{O}{iii} & 4.85 & 1.06 & 1.01 & 4.81 & 22.0 & 0.52 & 4.83 & 54.4 & 0.60 & - & - &  -\\        
        1416.93 & \ion{S}{iv} & 4.88 & 0.33 & 0.72 & 4.87 & 0.83 & 1.16 & 4.94 & 7.09 & 0.23 & 4.86 & 0.05 & 0.62 \\
        1486.50 & \ion{N}{iv} & 5.01 & 0.61 & 0.69 & 5.02 & 4.99 & 1.04 & 4.96 & 32.8 & 0.17 & 5.03 & 0.08 & 0.52 \\
        1406.06 & \ion{S}{iv} & 5.03 & 0.42 & 1.26 & 4.99 & 1.47 & 1.05 & - & - & - & 5.20 & 0.03 & 2.41 \\
        1404.78 & \ion{O}{iv} & 5.04 & 0.75 & 0.64 & 5.08 & 4.10 & 0.95 & 5.00 & 23.1 & 0.46 & 5.06 & 0.07 & 0.70 \\
        1399.77 & \ion{O}{iv} & 5.06 & 0.06 & 1.02 & 5.07 & 1.52 & 1.02 & 5.01 & 13.8 & 1.01 & 5.09 & 0.06 & 0.77 \\
        1407.39 & \ion{O}{iv} & 5.06 & 0.58 & 0.94 & 5.11 & 1.43 & 1.05 & 5.04 & 13.9 & 0.98 & 5.09 & 0.05 & 1.06 \\
        1401.16 & \ion{O}{iv} & 5.07 & 1.82 & 0.98 & 5.09 & 9.02 & 0.90 & 5.02 & 48.0 & 0.90 & 5.12 & 0.15 & 1.05 \\
        1199.18 & \ion{S}{v} & 5.09 & 0.47 & 1.34 & 5.11 & 1.64 & 1.04 & 5.05 & 7.42 & 1.24 & 5.15 & 0.04 & 1.52 \\
        1548.20 & \ion{C}{iv} & 5.18 & 50.1 & 1.33 & 5.15 & 188 & 0.91 & 5.34 & 668 & 1.20 & 5.44 & 8.95 & 0.82 \\
        1550.77 & \ion{C}{iv} & 5.18 & 26.8 & 1.24 & 5.37 & 103 & 0.84 & 5.34 & 366 & 1.10 & 5.44 & 4.93 & 0.75 \\
        1218.35 & \ion{O}{v} & 5.43 & 52.7 & 1.04 & 5.37 & 20.4 & 0.76 & 5.51 & 39.5 & 1.21 & 5.49 & 0.99 & 1.07\\
        1371.29 & \ion{O}{v} & 5.48 & 1.12 & 0.85 & 5.43 & 1.60 & 1.26 & 5.61 & 11.3 & 0.76 & 5.53 & 0.20 & 1.03 \\
        1238.82 & \ion{N}{v} & 5.73 & 8.68 & 1.19 & 5.69 & 24.3 & 0.81 & 5.87 & 103 & 1.17 & 5.82 & 2.87 & 1.07 \\
        1242.80 & \ion{N}{v} & 5.73 & 4.35 & 1.19 & 5.69 & 13.6 & 0.90 & 5.87 & 55.6 & 1.08 & 5.82 & 1.49 & 1.03 \\ 
        \hline
        \hline
    \end{tabular}
    \label{tab:all_results}
\end{table*}

\begin{figure*}
    \centerline{{\includegraphics[angle=90,width=9cm,keepaspectratio]{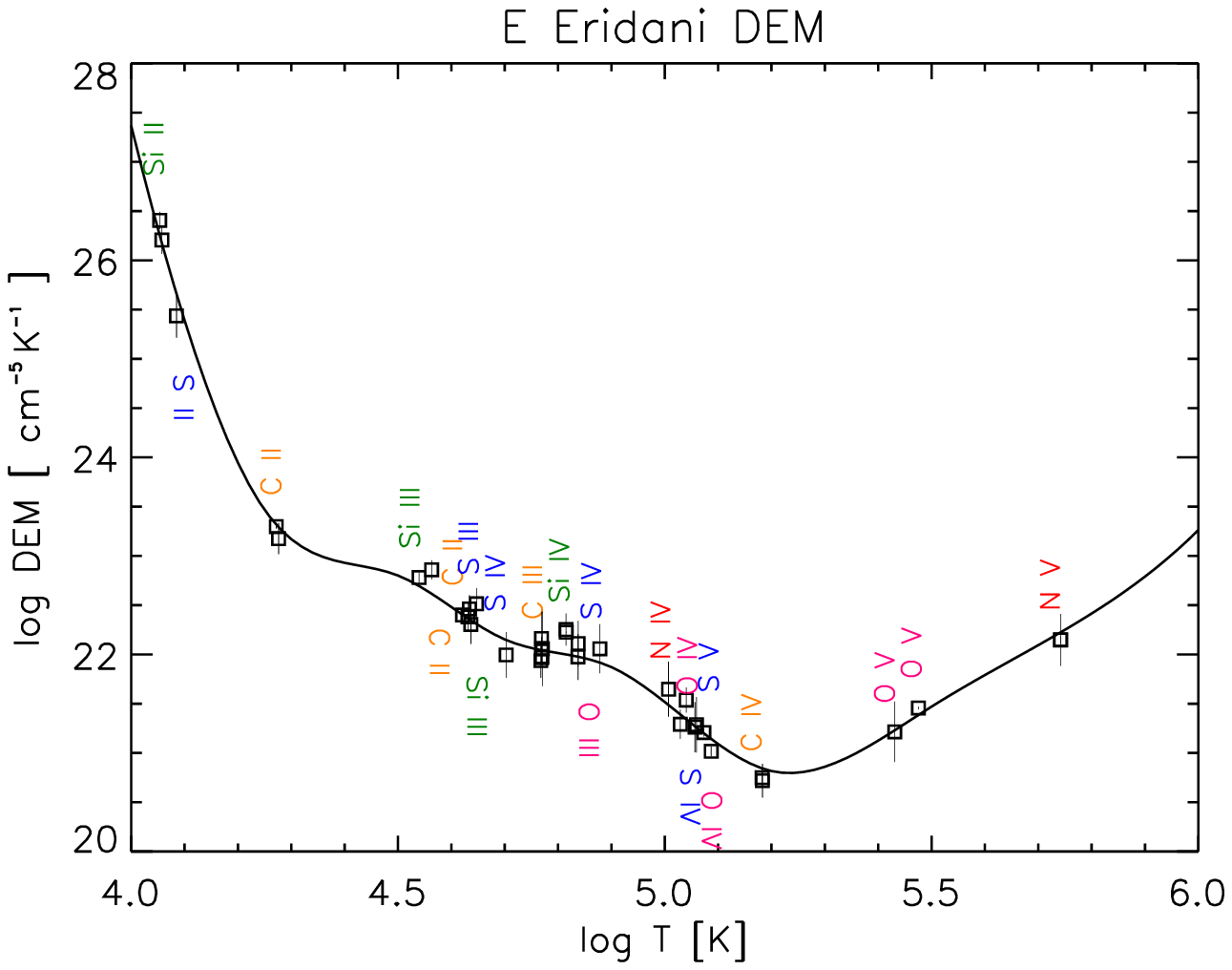}
    \includegraphics[angle=90, width=9cm, keepaspectratio]{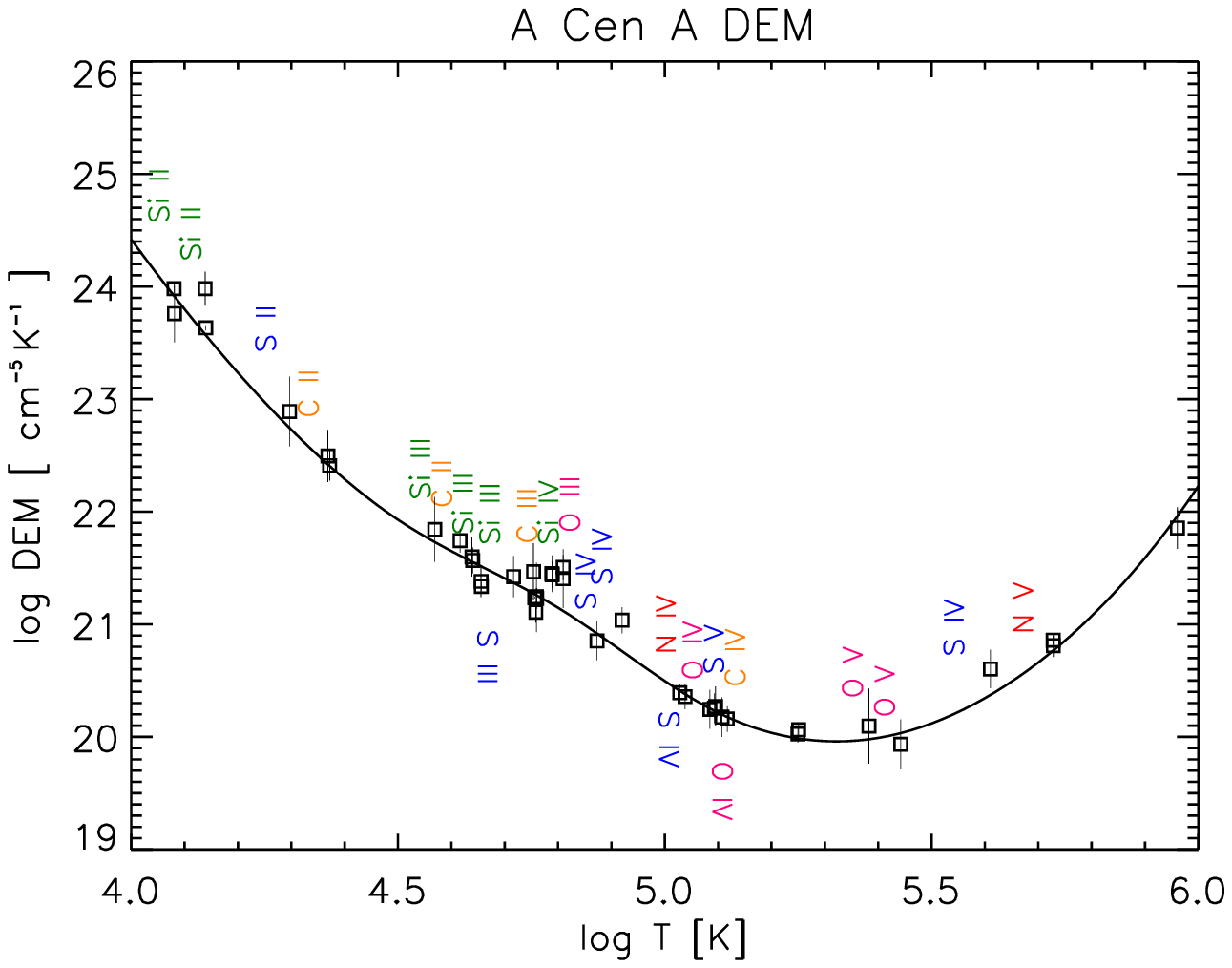}}}

    \centerline{{\includegraphics[angle=90,width=9cm,keepaspectratio]{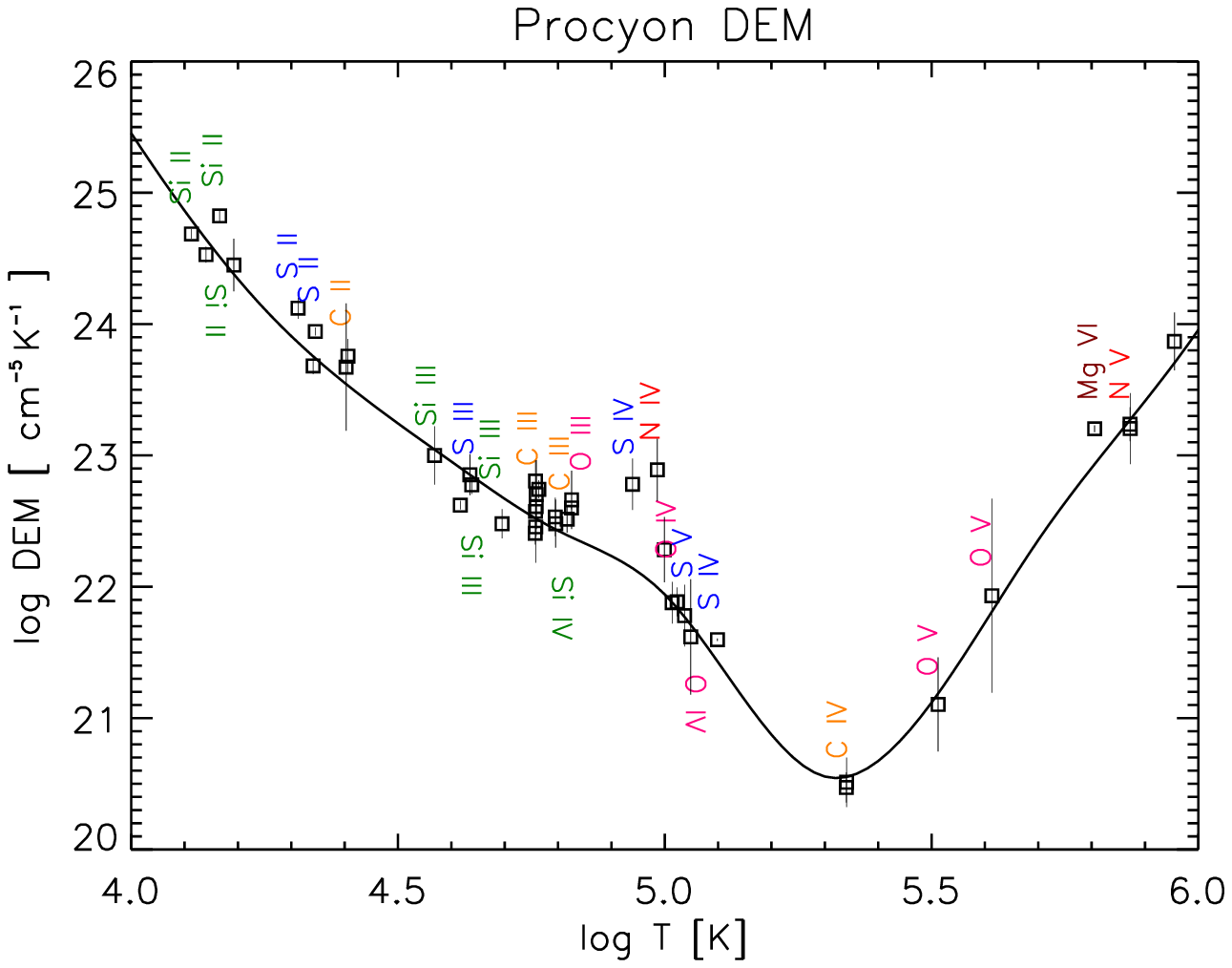}
    \includegraphics[angle=90, width=9cm, keepaspectratio]{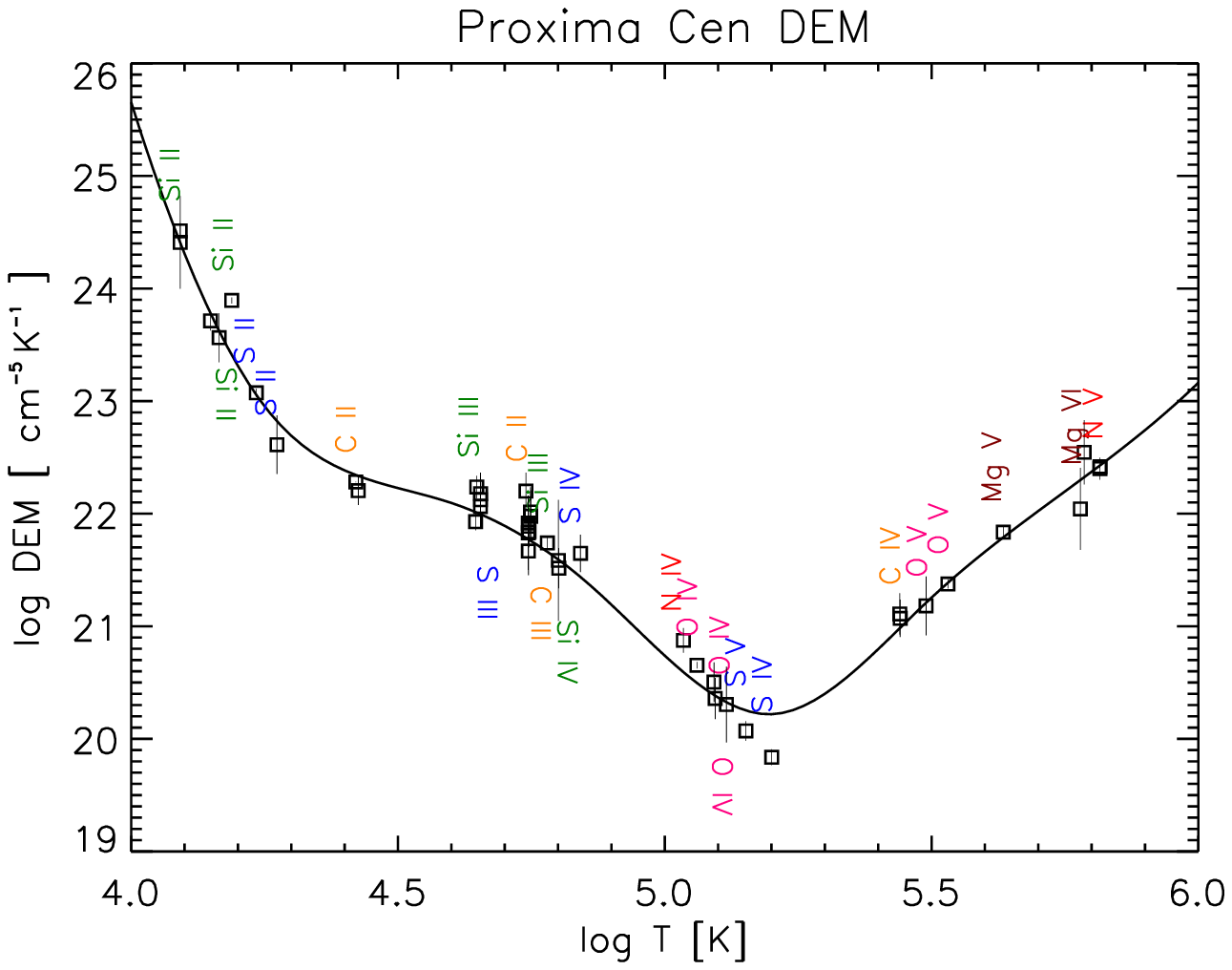}}}
    
    \caption{ DEM plots for $\epsilon$ Eridani, A Cen A, Procyon, and Proxima Centauri. Different ions are labelled and coloured according to their element.}
    \label{fig:DEM}
\end{figure*}

After fitting the STIS lines we commenced modelling the DEM for each star. Initially, we examined the contribution functions and removed lines influenced by noise, those that were very weak or those whose atomic data are lacking from the literature. For each star, we carefully reviewed the available photospheric abundances as discussed in Section~\ref{Stellar Sample}. Despite selecting the best available photospheric values, we continued refining the DEM fit, to also be able to accurately determine the presence of the FIP effect in these stellar TRs. 

\begin{figure*}
    \centering
    \includegraphics[width=0.7\textwidth]{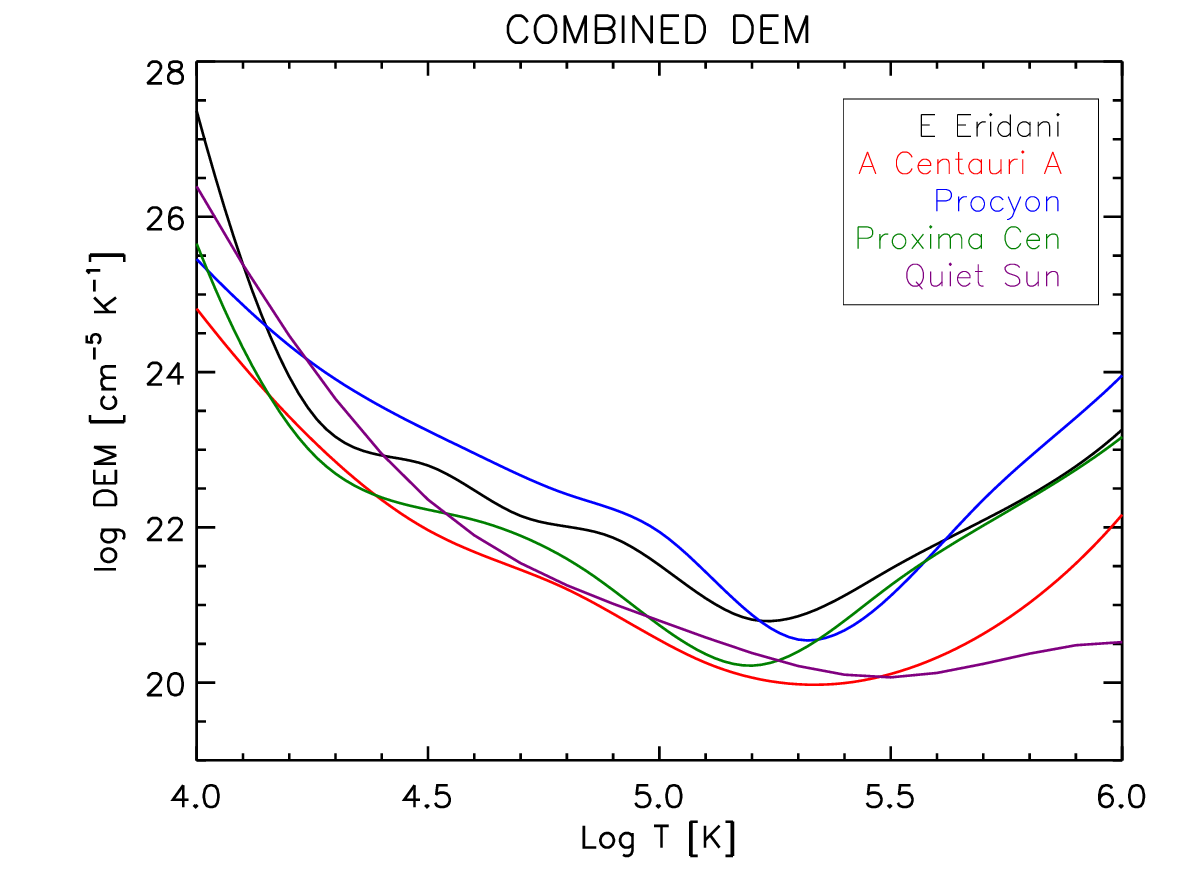}
    \caption{Combined DEM plots for $\epsilon$ Eridani, A Cen A, Procyon, Proxima Centauri, and the quiet Sun.}
    \label{fig:combined_dem}
\end{figure*}

We present our numerical results from the DEM analysis in Table~\ref{tab:all_results}, which lists the \chianti wavelengths and respective ion identifications for the main lines we used for the DEM analysis, most of which are either anomalous ions, strong emission lines, or significantly influencing the DEM. Although we fit many lines for each star, we provide a representative sample based on the aforementioned criteria. This table includes data for all four stars to facilitate comparison. For each star, we report three key parameters: the formation temperature of each ion, the observed fluxes derived from line fitting, and the ratio of calculated/predicted to observed flux. Following extensive fitting and modelling, we present the best DEM fits for each star in Figure~\ref{fig:DEM}, and a combined DEM plot in Figure~\ref{fig:combined_dem} to allow comparison between the stars and the quiet Sun.

In fig~\ref{fig:DEM} we observed a minimum in the DEM at log $T\approx$ 5.2 K for all stars (except Procyon, whose minimum point is found at log $T\approx$ 5.4 K). This is commonly observed, and is related to the radiative losses having a peak in the transition region (TR) emission, as outlined e.g. by \citet{jordan_chromospheres_1987}.

In fig~\ref{fig:combined_dem} we see a comparison of the DEM of all stars in the sample and an old quiet Sun DEM obtained from Skylab,
available within the CHIANTI database, derived using the \cite{vernazza_extreme_1978} average Quiet Sun and a constant pressure of $P_{\rm e}=$ 3 $\times 10^{15}$ cm$^{-3}$ K.
$\epsilon$ Eridani, shown in black, is an active star with a higher level of magnetic activity compared to the Sun, hence it is reasonable to appear higher in the log DEM axis, also suggesting an active corona with substantial high-temperature plasma. A Centauri A, shown in red, is a solar-like star but slightly older and less active, so the DEM is even lower but pretty close to the Sun and much lower at high temperatures compared to $\epsilon$ Eridani, again as expected. When it comes to Procyon which is an F-type star, shown in blue, as described earlier we would expect a quiet behaviour, however we see that although the DEM at low temperatures is much lower than the quiet Sun, it remains much higher than all of the stars in the TR, photosphere and chromosphere. This wasn't an expected behaviour, however we also found from the observed irradiances that its pressure is much higher than expected ($10^{16}$) which explains this DEM behaviour. Proxima Centauri is a red dwarf known for its flares, shown in green and shows a similar behaviour to Alpha Centauri A. Last but not least, the DEM of the quiet sun, shown in purple, is relatively low across all temperatures, reflecting less plasma emission compared to the active stars, as expected. It is evident that the Sun's TR can be described by photospheric abundances, as shown in \citet{dufresne_chianti_2024}. 

In general, the DEM is expected to rise after the minimum in temperature which indicates active regions, flares, or other forms of coronal heating that contribute to the high-temperature emission. 
The Quiet Sun and A Centauri A have a slow rise after the minimum, while Procyon, Proxima Centauri and $\epsilon$ Eridani rise significantly,
although observations from other wavelengths need to be
considered to constrain the upper TR and the coronae
of these stars.

There is plenty of evidence in the solar case that 
in remote-sensing observations the variations of the S abundance
follow those of the high-FIP elements such as Ar, despite
this element having FIP $=10$ eV \citep{del_zanna_solar_2018}. 
The solar S abundance, however, has been uncertain. 

For Alpha  Cen A we encountered small problems when fitting the lines at 1371~\AA\ (\ion{O}{V}), 1402~\AA\ (\ion{Si}{IV}), 1238~\AA\ and 1242~\AA\ (\ion{N}{V}), and 1501~\AA\ (\ion{Si}{III}). Despite these issues, the overall fitting was consistent with fluxes reported in other studies \citep{ayres_far-ultraviolet_2015, judge_understanding_2005, pagano_hststis_2004}, with a discrepancy of only 10$\%$.
Notably, we found no evidence of the FIP effect in the TR of A Cen A, contrary to the findings of \citet{raassen_chandra-letgs_2003}, which suggested its presence in the stellar corona.

For Procyon, the observational data for \ion{N}{IV} 1486~\AA\ and \ion{S}{IV} were very poor due to significant noise. Therefore, we excluded these lines from the DEM analysis. Similarly to the A Cen A findings, we found that there is no FIP effect in the TR of Procyon. Although Procyon is known as a quiet star, we found an increased DEM (see fig~\ref{fig:combined_dem}) compared to the rest of the sample, obtained with solar photospheric values.  

For Proxima Centauri, we did not find any published results in the literature for modelled UV fluxes to compare with our results. In addition, using the same abundances with A Cen A we found no FIP effect in TR of Proxima Centauri. 

For $\epsilon$ Eridani, we found that the advanced models employed in our study \citep{dufresne_chianti_2024} provide improved calculated-observed flux ratios for some ions, compared to the results obtained with the ionisation balance calculations used by \citet{sim_modelling_2005}. Regarding the FIP effect, our findings indicate an absence of this effect in this star's TR. 

\begin{table*}
    \centering
    \caption{Comparison table using $\epsilon$ Eridani as an example star from the sample. We present three cases: (1) Results from the advanced models in \textsc{chianti} v.11 under "Advanced", (2) results using the coronal approximation under "Coronal", and (3) results from \citet{sim_modelling_2005}. The table includes $F_{\text{obs}}$ (measured flux in $10^{-14}$ erg cm$^{-2}$ s$^{-1}$) for both our work and \citet{sim_modelling_2005}, $T_{\text{p}}$ [K] of the ions for the advanced models and coronal approximation, and $F_{\text{calc}}/F_{\text{obs}}$ for all three cases. "M1" and "M2" under Sim \& Jordan 2005 represent the two models we're comparing with, as explained in the text.}
    \begin{tabular}{c|c|cc|cc|cc|cc}
        \hline
        \hline
        \textbf{Wavelength (\AA)} & \textbf{Ion} & \multicolumn{2}{c|}{\textbf{$F_{\text{obs}}$}} & \multicolumn{2}{c|}{\textbf{Advanced}} & \multicolumn{2}{c|}{\textbf{Coronal}} & \multicolumn{2}{c}{\textbf{Sim \& Jordan 2005}} \\
        & & This work & Sim \& Jordan & \textbf{$\log(T_{\text{p}})$} & \textbf{$F_{\text{calc}}/F_{\text{obs}}$} & \textbf{$\log(T_{\text{p}})$} & \textbf{$F_{\text{calc}}/F_{\text{obs}}$} & $\frac{F_{\text{calc}}}{F_{\text{obs}}}$ (M1) & $\frac{F_{\text{calc}}}{F_{\text{obs}}}$ (M2) \\ 
        \hline
        1298.95 & \ion{Si}{iii} & 1.67 & 1.67 & 4.64 & 0.99 & 4.74 & 1.64 & 0.78 & 0.80 \\
        1294.54 & \ion{Si}{iii} & 0.59 & 0.52 & 4.64 & 0.80 & 4.74 & 1.33 & - & 0.73\\
        1402.77 & \ion{Si}{iv} & 10.6 & 12.1 & 4.82 & 0.63 & 4.92 & 0.19 & 0.71 & 0.64 \\
        1393.76 & \ion{Si}{iv} & 19.9 & 23 & 4.82 & 0.67 & 4.92 & 0.20 & - & 0.66\\
        1416.93 & \ion{S}{iv} & 0.33 & 0.42 & 4.88 & 0.83 & 4.95 & 0.62 & - & - \\
        1486.50 & \ion{N}{iv} & 0.61 & 0.45 & 4.99 & 0.74 & 5.14 & 0.47 & 1.30 & 1.32 \\
        1404.78 & \ion{O}{iv} & 0.75 & 0.69 & 5.03 & 0.68 & 5.20 & 0.59 & - & -\\
        1399.77 & \ion{O}{iv} & 0.56 & 0.54 & 5.05 & 1.07 & 5.21 & 0.83 & - & - \\
        1407.39 & \ion{O}{iv} & 0.58 & 0.52 & 5.06 & 0.98 & 5.21 & 0.76 & - & -\\
        1401.16 & \ion{O}{iv} & 1.82 & 1.81 & 5.07 & 1.03 & 5.23 & 0.88 & 1.89 & -\\
        1548.20 & \ion{C}{iv} & 50.1 & 55.9 & 5.14 & 1.33 & 5.40 & 0.39 & - & 0.68 \\
        1550.77 & \ion{C}{iv} & 26.8 & 28.2 & 5.14 & 1.24 & 5.40 & 0.42 & 0.81 & 0.68 \\
        1218.35 & \ion{O}{v} & 5.27 & 6.8 & 5.40 & 1.03 & 5.46 & 1.36 & 0.78 & - \\
        1371.29 & \ion{O}{v} & 1.12 & 1.05 & 5.44 & 0.81 & 5.49 & 1.19 & 0.60 & - \\
        1238.82 & \ion{N}{v} & 8.68 & 9.4 & 5.73 & 1.04 & 5.79 & 0.86 & 0.71 & - \\
        1242.80 & \ion{N}{v} & 4.35 & 4.6 & 5.73 & 1.04 & 5.79 & 0.86 & 0.71 & - \\
        \hline
        \hline
    \end{tabular}
    \label{tab:ratio_comparison}
\end{table*}

In Table~\ref{tab:ratio_comparison}, we compare the results 
we obtained with the advanced ionisation equilibrium models
with those with the zero-density coronal approximation for 
$\epsilon$ Eri as an example of the improvement. The Table also
includes a comparison with the results from \citet{sim_modelling_2005}. They present four different models in their paper, but we specifically compare our results with those derived from their simple EMD model (see "M1" in Table~\ref{tab:ratio_comparison} below) and their DEMD-A model (see "M2" in Table~\ref{tab:ratio_comparison}  below), which includes radiative transfer calculations. 

In table~\ref{tab:ratio_comparison}, we present their observed fluxes ($F_{\text{obs}}$) for a few ions, taken from \citet{sim_modelling_2005} and their $F_{\text{calc}}/F_{\text{obs}}$ ratio. We note that in most cases, the measured fluxes 
agree within 10\%. For several ions, \citet{sim_modelling_2005} 
only provide in their Table 7 their calculated flux
ratio for the multiplet, which we report in Table~\ref{tab:ratio_comparison}.

In addition, we note the comparison of the EMDs from the different models that \citet{sim_modelling_2005} present in their Figure 7. Although they conclude that their DEMD-B is the most favoured, Figure 7 reveals that the EMD of the DEMD-B model is not as smooth as the EMD from the other models that do not show significant improvements in the results.

We measured the stellar TR abundances, using the results of the DEM analysis. We started as a baseline with the stellar photospheric 
abundances, and checked if departures were 
required when doing the fitting. We present our estimates for the stellar TR abundances in Table~\ref{tab:TR stellar abundances}. These estimates were based on groups of ions forming at similar temperatures, ensuring that the ratios of calculated to observed fluxes were consistently accurate. As discussed in the earlier sections, these estimates were derived after reviewing photospheric abundances from the literature which we scaled to solar photospheric values by \cite{asplund_chemical_2021} and modelled the DEM using the advanced models. For Proxima Centauri, A Centauri A and Procyon we found complete consistency of photospheric values in the TR. For $\epsilon$ Eridani, we slightly adjusted the photospheric measurements (keeping C fixed) and we provide our adjustments in Table~\ref{tab:TR stellar abundances} while we also state the original photospheric values by \cite{zhao_chemical_2002} in parenthesis. However, we still see consistency of photospheric values in the TR of $\epsilon$ Eridani, allowing us to conclude that we find no FIP effect in any of the four stellar TR.

\begin{table}
    \centering
    \caption{Final estimates of stellar TR abundances for the stellar sample. For $\epsilon$ Eridani we provide our adjusted TR estimates while the original photospheric values by \citet{zhao_chemical_2002} are in parenthesis. For A Centauri A and Proxima Centauri, the values are taken from \citet{morel_chemical_2018}, and for Procyon we used values from \citet{drake_stellar_1995}.}
    \begin{tabular}{c|c|c|c|c}
        \hline
        \hline
        & $\epsilon$ Eridani & $\alpha$ Centauri A & Procyon & Proxima Centauri \\
        \hline
     {[Si/C]} & 0.10 (0.09) & 0.12 & 0.10 & 0.12 \\
    {[S/C]}  & 0.05 (0.05) & 0.05 & 0.05 & 0.05 \\
    {[S/Si]} & 0.45 (0.58) & 0.41 & 0.46 & 0.41 \\
    {[S/N]}  & 0.21 (0.28) & 0.19 & 0.14 & 0.19 \\
    {[S/O]}  & 0.03 (0.04) & 0.03 & 0.02 & 0.03 \\
    {[N/C]}  & 0.22 (0.19) & 0.25 & 0.31 & 0.25 \\
        \hline
        \hline
    \end{tabular}
    \label{tab:TR stellar abundances}
\end{table}

\section{Summary and Conclusions}
\label{Discussion}

We employed our advanced ionisation equilibrium models \citep{dufresne_chianti_2024} to improve the modelling of four FGKM active stars, A Cen A, Procyon, Proxima Centauri and $\epsilon$ Eridani. We significantly improved the calculated to observed flux ratios 
(see Table~\ref{tab:all_results} and Table~\ref{tab:ratio_comparison}),
especially for the anomalous ions, although there are exceptions such as Si IV, which is still under-predicted by up to 40\% in one case. 
We compared our results to those of \citet{sim_modelling_2005}, see Table~\ref{tab:ratio_comparison}, finding similar results for some ions,
but very different for others. 

 Further improvements could be obtained by e.g. changing the atmospheric model and the amount of neutral hydrogen,  as the formation temperature
of Si ions is most sensitive to charge transfer effects
\citep[cf][ and references therein]{dufresne_chianti_2024}.

We evaluated the improvements via a DEM analysis (see fig~\ref{fig:DEM}) and compared the results with the quiet Sun (see fig~\ref{fig:combined_dem}). We found that $\epsilon$ Eridani and Procyon show higher DEM values at high temperatures compared to A Cen A, Proxima Centauri and the quiet Sun, indicating more active and hotter coronae which will be further studied in future papers. The quiet Sun and A Cen A have lower DEM values, showing less emissions at high temperatures, which is characteristic of stars with less magnetic activity. Proxima Centauri, despite being a red dwarf, shows a considerable amount of high-temperature plasma, which is unusual for stars of its type.

From the DEM analysis we also found that the TR of these four stars can well be described by photospheric abundances, contrary to 
published results on  their  coronae. 
For $\epsilon$ Eridani, we made a small adjustment  to the abundances, fixing the C value by \citet{zhao_chemical_2002}. Our findings are in agreement with our results of the solar TR, where we also found photospheric abundances.

It is somewhat surprising to see such good agreement between 
predicted and observed fluxes in TR lines, given the very simplistic 
1-D DEM modelling of a static atmosphere in ionization equilibrium,
which  clearly does not correspond to reality.  
However, regardless of how the plasma density
is distributed in height and spatial regions within  the atmospheres 
of the stars, ions that are formed at similar temperatures 
should have similar emission measures. 
Our study,  given that the advanced ionization equilibrium \citep{dufresne_chianti_2024} significantly improve the atomic modelling in the TR, indicates that we can use the UV lines to provide abundance estimates in the TR, as it has been commonly 
carried out for stellar coronae in the XUV. 
The advantage of this type of modelling is that it is direct and 
does not require radiative transfer calculations.

In the Sun, the FIP effect is prominently observed at high temperatures (above 1 MK) in active regions with strong magnetic field concentrations, while the quiet Sun corona has photospheric abundances. 
The solar TR has near photospheric abundances. 
The spatial extent of active regions is relatively small, 
and further a significant fraction of the solar TR is not physically
connected to the corona, as there are plenty of low-lying 
cool  structures in the quiet Sun and in active regions.
The TR emission, which is primarily in the UV, likely reflects processes that are distinct from those in the magnetically strong corona.
Therefore, extrapolating to solar-like stars, we were expecting to 
find from UV measurements 
photospheric abundances in their transition regions.
This is indeed what we found, comparing the few relative 
abundances we obtained with the few photospheric measurements we 
found in the literature.  Despite the variations in the 
reference solar photospheric values, the scatter of literature 
values and the uncertainties in our simple models, a FIP 
effect of a factor of 3--4 is clearly not present in our stellar
sample. Consequently, we didn't find any FIP dependence on the spectral type in the TR.

Solar irradiance variations reveal that, while activity cycles are pronounced in the X-rays and EUV, there is minimal variability in the UV and associated TR emission. This relative stability is attributed to the TR's lesser influence by large-scale magnetic events that dominate the corona. This has been discussed in the literature, for example by  \citet{ayres_far-ultraviolet_2015} who investigated stellar activity cycles in the solar-like A Cen A. Consequently, this stability 
indicates that the TR abundances are close to photospheric regardless 
of the activity cycle of the star.

Clearly, time-dependent ionization and flows can somewhat 
affect the  ion charge state distributions, so there is scope
for further improvements in those areas.
We also identified  ions that could be useful to measure the FIP
effect but for which improved atomic data are needed. 
We are working towards improving the models for singly ionised ions
which require inclusion of photo-ionization and recombination into 
Rydberg states, plus radiative transfer effects. 
Finally, we will extend the advanced models to 
stellar coronae. The effects would be smaller than those we have found 
for the TR, but should be taken into account. 
We will then revisit the analyses of stellar coronae  using XUV spectra
for the same  sample to further investigate the FIP effect in these stars.

\section*{Acknowledgements}

ED acknowledges support from STFC (UK) via a studentship (2882461 related to ST/Y509127/1). GDZ acknowledges support from STFC (UK) via the consolidated grants  to the atomic astrophysics group at DAMTP, University of Cambridge (ST/P000665/1. and ST/T000481/1).

Data were obtained from the Multimission Archive at the Space Telescope Science Institute (MAST). STScI is operated by the Association of Universities for Research in Astronomy, Inc., under NASA contract NAS5-26555. Support for MAST for non-HST data is provided by the NASA Office of Space Science via grant NAG5-7584 and by other grants and contracts.

\section*{Data Availability}

The data underlying are available on the MAST website: \url{https://mast.stsci.edu/search/ui/#/hst}. The atomic data are taken from the \chianti database: \url{https://www.chiantidatabase.org}.



\bibliographystyle{mnras}
\bibliography{paper} 

\begin{thebibliography}{}
\makeatletter
\relax
\def\mn@urlcharsother{\let\do\@makeother \do\$\do\&\do\#\do\^\do\_\do\%\do\~}
\def\mn@doi{\begingroup\mn@urlcharsother \@ifnextchar [ {\mn@doi@}
  {\mn@doi@[]}}
\def\mn@doi@[#1]#2{\def\@tempa{#1}\ifx\@tempa\@empty \href
  {http://dx.doi.org/#2} {doi:#2}\else \href {http://dx.doi.org/#2} {#1}\fi
  \endgroup}
\def\mn@eprint#1#2{\mn@eprint@#1:#2::\@nil}
\def\mn@eprint@arXiv#1{\href {http://arxiv.org/abs/#1} {{\tt arXiv:#1}}}
\def\mn@eprint@dblp#1{\href {http://dblp.uni-trier.de/rec/bibtex/#1.xml}
  {dblp:#1}}
\def\mn@eprint@#1:#2:#3:#4\@nil{\def\@tempa {#1}\def\@tempb {#2}\def\@tempc
  {#3}\ifx \@tempc \@empty \let \@tempc \@tempb \let \@tempb \@tempa \fi \ifx
  \@tempb \@empty \def\@tempb {arXiv}\fi \@ifundefined
  {mn@eprint@\@tempb}{\@tempb:\@tempc}{\expandafter \expandafter \csname
  mn@eprint@\@tempb\endcsname \expandafter{\@tempc}}}

\bibitem[\protect\citeauthoryear{Allende~Prieto, Barklem, Lambert  \&
  Cunha}{Allende~Prieto et~al.}{2004}]{allende_prieto_s4n_2004}
Allende~Prieto C.,  Barklem P.~S.,  Lambert D.~L.,   Cunha K.,  2004, \mn@doi
  [Astronomy and Astrophysics] {10.1051/0004-6361:20035801}, 420, 183

\bibitem[\protect\citeauthoryear{Anders \& Grevesse}{Anders \&
  Grevesse}{1989}]{anders_abundances_1989}
Anders E.,  Grevesse N.,  1989, \mn@doi [Geochimica et Cosmochimica Acta]
  {10.1016/0016-7037(89)90286-X}, 53, 197

\bibitem[\protect\citeauthoryear{Asplund, Amarsi  \& Grevesse}{Asplund
  et~al.}{2021}]{asplund_chemical_2021}
Asplund M.,  Amarsi A.~M.,   Grevesse N.,  2021, \mn@doi [Astronomy and
  Astrophysics] {10.1051/0004-6361/202140445}, 653, A141

\bibitem[\protect\citeauthoryear{Ayres}{Ayres}{2015}]{ayres_far-ultraviolet_2015}
Ayres T.~R.,  2015, \mn@doi [The Astronomical Journal]
  {10.1088/0004-6256/149/2/58}, 149, 58

\bibitem[\protect\citeauthoryear{Buder et~al.,}{Buder
  et~al.}{2021a}]{buder_galah_2021}
Buder S.,  et~al., 2021a, \mn@doi [Monthly Notices of the Royal Astronomical
  Society] {10.1093/mnras/stab1242}, 506, 150

\bibitem[\protect\citeauthoryear{Buder et~al.,}{Buder
  et~al.}{2021b}]{buder_vizier_2021}
Buder S.,  et~al., 2021b, VizieR Online Data Catalog, 750, J/MNRAS/506/150

\bibitem[\protect\citeauthoryear{Del~Zanna \& Andretta}{Del~Zanna \&
  Andretta}{2015}]{del_zanna_euv_2015}
Del~Zanna G.,  Andretta V.,  2015, \mn@doi [Astronomy \& Astrophysics]
  {10.1051/0004-6361/201526804}, 584, A29

\bibitem[\protect\citeauthoryear{Del~Zanna \& Mason}{Del~Zanna \&
  Mason}{2018}]{del_zanna_solar_2018}
Del~Zanna G.,  Mason H.~E.,  2018, \mn@doi [Living Reviews in Solar Physics]
  {10.1007/s41116-018-0015-3}, 15, 5

\bibitem[\protect\citeauthoryear{Del~Zanna, Landini  \& Mason}{Del~Zanna
  et~al.}{2002}]{del_zanna_spectroscopic_2002}
Del~Zanna G.,  Landini M.,   Mason H.~E.,  2002, \mn@doi [Astronomy and
  Astrophysics] {10.1051/0004-6361:20020164}, 385, 968

\bibitem[\protect\citeauthoryear{Del~Zanna, Dere, Young  \& Landi}{Del~Zanna
  et~al.}{2021}]{del_zanna_chiantiatomic_2021}
Del~Zanna G.,  Dere K.~P.,  Young P.~R.,   Landi E.,  2021, \mn@doi [The
  Astrophysical Journal] {10.3847/1538-4357/abd8ce}, 909, 38

\bibitem[\protect\citeauthoryear{Dere, Landi, Mason, Monsignori~Fossi  \&
  Young}{Dere et~al.}{1997}]{dere_chianti_1997}
Dere K.~P.,  Landi E.,  Mason H.~E.,  Monsignori~Fossi B.~C.,   Young P.~R.,
  1997, \mn@doi [Astronomy and Astrophysics Supplement Series]
  {10.1051/aas:1997368}, 125, 149

\bibitem[\protect\citeauthoryear{Drake, Laming  \& Widing}{Drake
  et~al.}{1995}]{drake_stellar_1995}
Drake J.~J.,  Laming J.~M.,   Widing K.~G.,  1995, \mn@doi [The Astrophysical
  Journal] {10.1086/175533}, 443, 393

\bibitem[\protect\citeauthoryear{Drake, Laming  \& Widing}{Drake
  et~al.}{1997}]{drake_stellar_1997}
Drake J.~J.,  Laming J.~M.,   Widing K.~G.,  1997, \mn@doi [The Astrophysical
  Journal] {10.1086/303755}, 478, 403

\bibitem[\protect\citeauthoryear{Dufresne, Del~Zanna  \& Badnell}{Dufresne
  et~al.}{2021}]{dufresne_influence_2021}
Dufresne R.~P.,  Del~Zanna G.,   Badnell N.~R.,  2021, \mn@doi [Monthly Notices
  of the Royal Astronomical Society] {10.1093/mnras/stab514}, 503, 1976

\bibitem[\protect\citeauthoryear{Dufresne, Del~Zanna  \& Mason}{Dufresne
  et~al.}{2023}]{dufresne_benchmark_2023}
Dufresne R.~P.,  Del~Zanna G.,   Mason H.~E.,  2023, \mn@doi [Monthly Notices
  of the Royal Astronomical Society] {10.1093/mnras/stad794}, 521, 4696

\bibitem[\protect\citeauthoryear{Dufresne, Del~Zanna, Young, Dere,
  Deliporanidou, Barnes  \& Landi}{Dufresne
  et~al.}{2024}]{dufresne_chianti_2024}
Dufresne R.~P.,  Del~Zanna G.,  Young P.~R.,  Dere K.~P.,  Deliporanidou E.,
  Barnes W.~T.,   Landi E.,  2024, {CHIANTI} -- an atomic database for emission
  lines -- {Paper} {XVIII}. {Version} 11, advanced ionization equilibrium
  models: density and charge transfer effects,
  \mn@doi{10.48550/arXiv.2403.16922}, \url {http://arxiv.org/abs/2403.16922}

\bibitem[\protect\citeauthoryear{Edvardsson, Andersen, Gustafsson, Lambert,
  Nissen  \& Tomkin}{Edvardsson et~al.}{1993}]{edvardsson_chemical_1993}
Edvardsson B.,  Andersen J.,  Gustafsson B.,  Lambert D.~L.,  Nissen P.~E.,
  Tomkin J.,  1993, Astronomy and Astrophysics Supplement Series, 102, 603

\bibitem[\protect\citeauthoryear{Fontenla, Landi, Snow  \& Woods}{Fontenla
  et~al.}{2014}]{fontenla_far-_2014}
Fontenla J.~M.,  Landi E.,  Snow M.,   Woods T.,  2014, \mn@doi [Solar Physics]
  {10.1007/s11207-013-0431-4}, 289, 515

\bibitem[\protect\citeauthoryear{Fuhrmeister, Lalitha, Poppenhaeger, Rudolf,
  Liefke, Reiners, Schmitt  \& Ness}{Fuhrmeister
  et~al.}{2011}]{fuhrmeister_multi-wavelength_2011}
Fuhrmeister B.,  Lalitha S.,  Poppenhaeger K.,  Rudolf N.,  Liefke C.,  Reiners
  A.,  Schmitt J. H. M.~M.,   Ness J.~U.,  2011, \mn@doi [Astronomy and
  Astrophysics] {10.1051/0004-6361/201117447}, 534, A133

\bibitem[\protect\citeauthoryear{Fuhrmeister et~al.,}{Fuhrmeister
  et~al.}{2022}]{fuhrmeister_astronomy_2022}
Fuhrmeister B.,  et~al., 2022, ] {10.1051/0004-6361/202243077}

\bibitem[\protect\citeauthoryear{Güdel, Audard, Reale, Skinner  \&
  Linsky}{Güdel et~al.}{2004}]{gudel_flares_2004}
Güdel M.,  Audard M.,  Reale F.,  Skinner S.~L.,   Linsky J.~L.,  2004,
  \mn@doi [Astronomy and Astrophysics] {10.1051/0004-6361:20031471}, 416, 713

\bibitem[\protect\citeauthoryear{Jordan, Ayres, Brown, Linsky  \& Simon}{Jordan
  et~al.}{1987}]{jordan_chromospheres_1987}
Jordan C.,  Ayres T.~R.,  Brown A.,  Linsky J.~L.,   Simon T.,  1987, \mn@doi
  [Monthly Notices of the Royal Astronomical Society]
  {10.1093/mnras/225.4.903}, 225, 903

\bibitem[\protect\citeauthoryear{Judge}{Judge}{2005}]{judge_understanding_2005}
Judge P.~G.,  2005, \mn@doi [Journal of Quantitative Spectroscopy and Radiative
  Transfer] {10.1016/j.jqsrt.2004.08.009}, 92, 479

\bibitem[\protect\citeauthoryear{Judge, Woods, Brekke  \& Rottman}{Judge
  et~al.}{1995}]{judge_failure_1995}
Judge P.~G.,  Woods T.~N.,  Brekke P.,   Rottman G.~J.,  1995

\bibitem[\protect\citeauthoryear{Judge, Saar, Carlsson, Ayres, Judge, Saar,
  Carlsson  \& Ayres}{Judge et~al.}{2004}]{judge_comparison_2004}
Judge P.~G.,  Saar S.~H.,  Carlsson M.,  Ayres T.~R.,  Judge P.~G.,  Saar
  S.~H.,  Carlsson M.,   Ayres T.~R.,  2004, \mn@doi [ApJ] {10.1086/421044},
  609, 392

\bibitem[\protect\citeauthoryear{Lalitha et~al.,}{Lalitha
  et~al.}{2020}]{lalitha_proxima_2020}
Lalitha S.,  et~al., 2020, \mn@doi [Monthly Notices of the Royal Astronomical
  Society] {10.1093/mnras/staa2574}, 498, 3658

\bibitem[\protect\citeauthoryear{Laming}{Laming}{2004}]{laming_unified_2004}
Laming J.~M.,  2004, \mn@doi [The Astrophysical Journal] {10.1086/423780}, 614,
  1063

\bibitem[\protect\citeauthoryear{Laming}{Laming}{2015}]{laming_fip_2015}
Laming J.~M.,  2015, \mn@doi [Living Reviews in Solar Physics]
  {10.1007/lrsp-2015-2}, 12, 2

\bibitem[\protect\citeauthoryear{Laming}{Laming}{2017}]{laming_first_2017}
Laming J.~M.,  2017, \mn@doi [The Astrophysical Journal]
  {10.3847/1538-4357/aa7cf1}, 844, 153

\bibitem[\protect\citeauthoryear{Laming, Drake  \& Widing}{Laming
  et~al.}{1995}]{laming_stellar_1995}
Laming J.~M.,  Drake J.~J.,   Widing K.~G.,  1995, \mn@doi [The Astrophysical
  Journal] {10.1086/175534}, 443, 416

\bibitem[\protect\citeauthoryear{Laming, Drake  \& Widing}{Laming
  et~al.}{1996}]{laming_stellar_1996}
Laming J.~M.,  Drake J.~J.,   Widing K.~G.,  1996, \mn@doi [The Astrophysical
  Journal] {10.1086/177208}, 462, 948

\bibitem[\protect\citeauthoryear{Linsky}{Linsky}{2019}]{linsky_host_2019}
Linsky J.,  2019, \mn@doi [Lecture Notes in Physics, Berlin Springer Verlag]
  {10.1007/978-3-030-11452-7}, 955

\bibitem[\protect\citeauthoryear{Morel}{Morel}{2018}]{morel_chemical_2018}
Morel T.,  2018, \mn@doi [Astronomy and Astrophysics]
  {10.1051/0004-6361/201833125}, 615, A172

\bibitem[\protect\citeauthoryear{Ness \& Jordan}{Ness \&
  Jordan}{2007}]{ness_corona_2007}
Ness J.-U.,  Jordan C.,  2007, \mn@doi [Monthly Notices of the Royal
  Astronomical Society] {10.1111/j.1365-2966.2007.12757.x}, 385, 1691

\bibitem[\protect\citeauthoryear{Pagano, Linsky, Valenti  \& Duncan}{Pagano
  et~al.}{2004}]{pagano_hststis_2004}
Pagano I.,  Linsky J.~L.,  Valenti J.,   Duncan D.~K.,  2004, \mn@doi
  [Astronomy and Astrophysics] {10.1051/0004-6361:20034002}, 415, 331

\bibitem[\protect\citeauthoryear{Pottasch, {Pottasch}  \& R}{Pottasch
  et~al.}{1963}]{pottasch_lower_1963}
Pottasch S.~R.,  {Pottasch}  R S.,  1963, \mn@doi [ApJ] {10.1086/147569}, 137,
  945

\bibitem[\protect\citeauthoryear{Raassen et~al.,}{Raassen
  et~al.}{2002}]{raassen_high-resolution_2002}
Raassen A. J.~J.,  et~al., 2002, \mn@doi [Astronomy \& Astrophysics]
  {10.1051/0004-6361:20020529}, 389, 228

\bibitem[\protect\citeauthoryear{Raassen, Ness, Mewe, van~der Meer, Burwitz  \&
  Kaastra}{Raassen et~al.}{2003}]{raassen_chandra-letgs_2003}
Raassen A. J.~J.,  Ness J.~U.,  Mewe R.,  van~der Meer R. L.~J.,  Burwitz V.,
  Kaastra J.~S.,  2003, \mn@doi [Astronomy and Astrophysics]
  {10.1051/0004-6361:20021899}, 400, 671

\bibitem[\protect\citeauthoryear{Rao, Del~Zanna, Mason  \& Dufresne}{Rao
  et~al.}{2022}]{rao_path-lengths_2022}
Rao Y.~K.,  Del~Zanna G.,  Mason H.~E.,   Dufresne R.,  2022, \mn@doi [Monthly
  Notices of the Royal Astronomical Society] {10.1093/mnras/stac2772}, 517,
  1422

\bibitem[\protect\citeauthoryear{Sanz-Forcada, Favata  \& Micela}{Sanz-Forcada
  et~al.}{2004}]{sanz-forcada_coronal_2004}
Sanz-Forcada J.,  Favata F.,   Micela G.,  2004, \mn@doi [Astronomy and
  Astrophysics] {10.1051/0004-6361:20034466}, 416, 281

\bibitem[\protect\citeauthoryear{Schmitt \& Ness}{Schmitt \&
  Ness}{2004}]{schmitt_coronal_2004}
Schmitt J. H. M.~M.,  Ness J.~U.,  2004, \mn@doi [Astronomy and Astrophysics]
  {10.1051/0004-6361:20031470}, 415, 1099

\bibitem[\protect\citeauthoryear{Seli et~al.,}{Seli
  et~al.}{2022}]{seli_extending_2022}
Seli B.,  et~al., 2022, \mn@doi [Astronomy \& Astrophysics]
  {10.1051/0004-6361/202141493}, 659, A3

\bibitem[\protect\citeauthoryear{Sim \& Jordan}{Sim \&
  Jordan}{2005}]{sim_modelling_2005}
Sim S.~A.,  Jordan C.,  2005, \mn@doi [MNRAS]
  {10.1111/j.1365-2966.2005.09247.x}, 361, 1102

\bibitem[\protect\citeauthoryear{Telleschi, Güdel, Briggs, Audard, Ness  \&
  Skinner}{Telleschi et~al.}{2005}]{telleschi_coronal_2005}
Telleschi A.,  Güdel M.,  Briggs K.,  Audard M.,  Ness J.-U.,   Skinner S.~L.,
   2005, \mn@doi [The Astrophysical Journal] {10.1086/428109}, 622, 653

\bibitem[\protect\citeauthoryear{Vernazza \& Reeves}{Vernazza \&
  Reeves}{1978}]{vernazza_extreme_1978}
Vernazza J.~E.,  Reeves E.~M.,  1978, \mn@doi [The Astrophysical Journal
  Supplement Series] {10.1086/190539}, 37, 485

\bibitem[\protect\citeauthoryear{Wood \& Linsky}{Wood \&
  Linsky}{2006}]{wood_coronal_2006}
Wood B.~E.,  Linsky J.~L.,  2006, \mn@doi [The Astrophysical Journal]
  {10.1086/501521}, 643, 444

\bibitem[\protect\citeauthoryear{Wood \& Linsky}{Wood \&
  Linsky}{2010}]{wood_resolving_2010}
Wood B.~E.,  Linsky J.~L.,  2010, \mn@doi [The Astrophysical Journal]
  {10.1088/0004-637X/717/2/1279}, 717, 1279

\bibitem[\protect\citeauthoryear{Zhao, Chen, Qiu  \& Li}{Zhao
  et~al.}{2002}]{zhao_chemical_2002}
Zhao G.,  Chen Y.~Q.,  Qiu H.~M.,   Li Z.~W.,  2002, \mn@doi [The Astronomical
  Journal] {10.1086/342862}, 124, 2224

\makeatother
\end{thebibliography}


\bsp	
\label{lastpage}
\end{document}